\documentclass[a4paper,11pt]{article}
\pdfoutput=1

\usepackage{jheppub}

\usepackage[T1]{fontenc}

\title{\boldmath From Special Relativity to embedded generators in Cartan subalgebras of rank-4 spin algebras}

\author[a,b,1]{Zen-chen Leon}

\affiliation[a]{The Affiliated High School of SCNU\\Guangzhou 510630, China}
\affiliation[b]{Guangdong Olympic School\\Guangzhou 510630, China}

\emailAdd{talie.firzen@foxmail.com}

\abstract{Starting by revisiting Special Relativity, here we provide a reliable characterization of the entire 4-dimensional fundamental structures in our reality where the frame of discrete tangent space of $F^{1,3}$ is quantized to massless, zero-momentum particles distributing on a 4-dimensional regular base $\{\mathbb{N}\cdot c_\alpha\}$ with metric $B(c_\alpha,c_\beta)=\delta_{\alpha\beta}=\text{diag}(+,+,+,+)$, determining the constant $c$ locally, as well as instant characterizations on all particles moving along the proper time of $\tau\in\mathbb{N}$. Together with $\phi_\text{IV}$ on 1-dimensional space $\{\mathbb{N}\cdot c_4\}$ of $B(c_\alpha,c_4)=0$, the quantized particles of tangent frame are split anti-symmetrically from roots $\gamma_{\alpha4}$ in a rank-4 Lie algebra with exactly $c_\alpha$ the generators of its Cartan subalgebra. As with the combined frame-Higgs valued in $\gamma_{\alpha4}$, every massive particle is related to some split frame and Higgs to obtain its unique quantized relative 3-velocity, 4-velocity, and rest mass under the restriction of Pauli Exclusion Principle at each $\tau$. We calculated the domains $S_i\subset F$ and $S_0\subset F_\geq$ in $F_\geq\oplus F^3$ the spacetime in which a massive particle $\mathtt{p}$ evolves with $\{x^\alpha\}$ of $x^\alpha\in S_\alpha$ the coordinate its Center-of-Mass under an eligible observation that $\vert F\vert=\vert F_\geq\vert=\aleph_0<\vert\mathbb{R}\vert$, available for post-Newtonian approaches.}

\begin{document} 
\maketitle
\flushbottom

\section{Introduction}
\label{sec:intro}

The 4-dimensional spacetime with metric $\eta_{\alpha\beta}=\text{diag}(+,-,-,-)$ in our reality was discovered by A. Einstein in 1905.

In light of an invariant form of Maxwell's equations, the restriction of $(d\tau)^2=(dx^0)^2-\sum_i(dx^i)^2$ within the well-ordered proper time $\tau$ was derived in Special Relativity (SR), in which all of the considered massive particles with Center-of-Mass (CM), and massless particles with invariant 3-velocity $c$ are evolving. SR is eligible to be considered as the foundation of both continuous and discrete approaches in elementary gauge theories. On a continuous aspect, we work General Relativity (GR) \cite{1,2,3,4,5,6,7,8,9,10} with Energy-Momentum Tensor (EMT) \cite{11,12,13,14,15,16,17,18,19,20}, Einstein-Cartan gravity \cite{21,22,23,24,25,26,27,28,29,30}, MacDowell-Mansouri gravity \cite{31,32,33}, Chern-Simons (CS) gravity \cite{34,35,36}, Loop Quantum Gravity (LQG) \cite{37,38,39,40,41,42}, or other methods as teleparallel $f(T)$ gravity \cite{43,44,45,46,47,48,49} on gravitational sections. As well as perturbative Quantum Field Theory (QFT) on Quantum Electrodynamics (QED), Quantum Chromodynamics (QCD), and weak interactions in the Standard Model (SM), we described the elementary particles as smoothly continuous, Lie-valued functional of fields over the conformal base manifold of 4-dimensional spacetime. Under a Lagrangian representation, all of the equations of motion of diverse sections of fields as Klein-Gordon equations, Dirac equations, and Lorentz gauged, bosonic equations on vectors or tensors \cite{50,51,52}, are derived from basic regulations in SR as $\langle E\rangle^2=m^2c^4+\langle p\rangle^2c^2$ with additional arrangements, as 4-component Dirac fermions are quantized in Schrodinger equation in Quantum Mechanics (QM) to leave a first-ordered kinetic term in its Lagrangian \cite{53}. In this case, connected Green functions of Schwinger sources and Feynman propagators are generally introduced to find the contribution of all vertices in Feynman diagrams of a specific interaction raised from a non-linear interacting term in the considered Lagrangian, before necessarily determining all of the cut-off scales in renormalization groups (RNG). Other descriptions or Grand Unified Theories (GUTs) beyond the SM in this way may focus more on algebraic selections of gauged Lie groups, but we cannot ignore the renormalized perturbations in QFT \cite{54,55,56}, ever since we have considered the particles as continuous sources.

However, the great amount of computation in perturbative QFT seems to make itself inefficient and in a sense, it is inaccurate because currently we can only calculate on finite levels of Feynman diagrams. Partly because of this, attempts on post-Newtonian electromagnetism, gravity, and spin precession were made \cite{57,58,59,60,61,62}. In this case, massive particles with CM in their reference frames are restricted by another series of equations of motion \cite{63,64,65,66,67,68,69,70}. A discrete frame can be introduced to simplify our calculations and makes our results accurate, but it is hard and necessary to cover all the properties with certain structures and discrete functional over the space. Therefore, if we devote ourselves to discrete developments (but not approximative simulations as lattice gauge theories), we need firstly to revisit SR, the evolution of elementary particles.

Here I want to present the entire internal structures of the 4-dimensional spacetime generated by $\gamma_\alpha\in Cl^1(1,3)$ with metric $\eta_{\alpha\beta}=\text{diag}(+,-,-,-)$, while the quantized frame of its quantized tangent space is discretely embedded on a 4-dimensional Cartan subalgebra $\{\mathbb{N}\cdot c_\alpha\}$ with metric $\delta_{\alpha\beta}=\text{diag}(+,+,+,+)$.

In this paper, the indices $\alpha,\beta=0,1,2,3$, $\mu,\nu=0,1,2,3$, and $i,j=1,2,3$ are arranged in Einstein summation convention. Latin letters in parentheses as $(i)$, $i\in\mathbb{N}$ or $(q)$, $q\in\mathbb{N}$ label different considered particles, e.g. $\mathtt{p}_{(1)}\neq\mathtt{p}_{(2)}$.

\section{Kinematic SR on massive particles}

We start our investigation from revisiting the basic regulations on massive particles moving along the proper time $\tau$ in SR. As well as considered in Newtonian classical mechanics, a massive particle (or a mass point) $\mathtt{p}$ refers to a number $m\in S_4\subset\mathbb{R}_+$ of its \textit{mass}, and a set of numbers $\{x^\alpha\}$, $\alpha=0, 1, 2, 3$ of the coordinates of its Center-of-Mass (CM), valued in $\forall\ \alpha: x^\alpha\in S_\alpha\subset\mathbb{R}$ where the domains $S_\alpha$ and $S_4$ can be either continuous or discrete. The \textit{CM-coordinate} $\{x^{\alpha(i)}\}$ and $\{x^{\alpha(j)}\}$, and masses $m^{(i)}$ and $m^{(j)}$ of two different massive particles $\mathtt{p}_{(i)}$ and $\mathtt{p}_{(j)}$, $i\neq j$ distinguished by labels $i,j\in \mathbb{Z_+}$ sit inside the same domains $S_\alpha$ and $S_4$, while their difference are presented by $m^{(i-j)}\equiv m^{(j)}-m^{(i)}$ and $x^{\alpha(i-j)}\equiv x^{\alpha(j)}-x^{\alpha(i)}$ that
\begin{equation}
i\neq j\ \ \ \Leftarrow\ \ \ m^{(i-j)}\neq0\ \ \ \text{or}\ \ \ x^{\alpha(i-j)}\neq0
\end{equation}

It is a truth that every massive particle moves within a proper time $\tau\in F$, where the linearly ordered set $F$ be either continuous or discrete as $F=\mathbb{R},\ \mathbb{Z},\ \mathbb{R}\verb"\"\mathbb{R}_-,\ \text{or}\ \mathbb{N}$. At each $\tau$, the observed CM of a massive particle has its coordinates $\{x^\alpha(\tau)\}\in\{S_\alpha\}$ which may vary to
\begin{equation}
d\tau\equiv\tau'-\tau>0\ \ \ \ \ \ dx^\alpha\equiv x^\alpha(\tau')-x^\alpha(\tau)
\end{equation}
and in this case, its mass in its rest frame always remain invariant. The constraint discovered on every observed massive particles in SR is (now in natural units of $c=1$)
\begin{equation}
\begin{split}
(d\tau)^2 &= \eta_{\alpha\beta}dx^\alpha dx^\beta\equiv dx^\alpha dx_\alpha\\
&= dx^0dx_0+dx^1dx_1+dx^2dx_2+dx^3dx_3\\
&= (dx^0)^2-(dx^1)^2-(dx^2)^2-(dx^3)^2>0
\end{split}
\end{equation}
where the metric $\eta_{\alpha\beta}=\text{diag}(+,-,-,-)$ is declared from this constraint and $dx_\alpha\equiv\eta_{\alpha\beta}dx^\beta$ is defined by a contraction on index $\beta$. It is different from a 3D Newtonian constraint, $(d\tau_0)^2\equiv dx^idx_i=dx^1dx_1+dx^2dx_2+dx^3dx_3$ with the metric $\delta_{ij}=\text{diag}(+,+,+)$ that $x_i\equiv\delta_{ij}x^j=x^i$, and no observation is required to determine the $dx^\alpha$. Commonly, Eq.$(2.3)$ holds for every $d\tau\geq0$ and on a differential level, it holds for $d\tau\rightarrow0^+$ when $\tau\in F$ is continuous, and for $\nexists\ \tau'': \tau'>\tau''>\tau$ when $\tau\in F$ is discrete. 

Now we could find the observed rates $\beta^i\equiv\dfrac{dx^i}{dx^0}$ of $\beta^i\equiv\beta^{i(2-1)}$ (similarly hereinafter) instantly, to present the motion of a specific massive particle $\mathtt{p}_{(1)}$. If we define $\beta\equiv\sqrt{\beta^i\beta_i}$, then its range is $\beta\in(0,1)$ for $d\tau>0$. In this range, the rates $V^\alpha\equiv\dfrac{dx^\alpha}{d\tau}$ and $V_\alpha\equiv\eta_{\alpha\beta}V^\beta$ satisfying $V^\alpha V_\alpha=1$ can be transformed via $\beta^i=\dfrac{V^i}{V^0}$, i.e. $V^i=V^0\beta^i$ with ordinary multiplication between numbers $V^0$ and $\beta^i$, where $V^0=\dfrac{1}{\sqrt{1-\beta^2}}\equiv\gamma$ derived from Eq.$(2.3)$ is currently dimensionless. Here we have not introduced the entire quadratic space of SR in which $\{x^\alpha\}$ are varying, so temporarily $\beta^i$ and $V^\alpha$ are real numbers, standing for the norms of vector components of observed (or relevant) 3-velocity $\overrightarrow{\beta}$ and 4-velocity $\overrightarrow{V}$.

When we find the norms of 3-momentum components as $p^i\equiv m\beta^i$ for some $\tau$ and $\mathtt{p}_{(1)}$, the range of $p\equiv\sqrt{p^ip_i}$ is $p\in[0,m)$. In this range, the norm of its 4-momentum components $P^\alpha\equiv mV^\alpha$ and $P_\alpha\equiv\eta_{\alpha\beta}P^\beta=\eta_{\alpha\beta}mV^\beta$ satisfy $P^\alpha P_\alpha=m^2$.

Only by observing $\mathtt{p}_{(1)}$ from another massive particle $\mathtt{p}_{(2)}$ as the instant (and inertial) observer, could there be a \textit{1+3 decomposition} \cite{71,72} of $\mathtt{p}_{(1)}$ as
\begin{equation}
\begin{cases}
P^{0(1)}+P^{i(1)}=\langle E\rangle(V^{0(2)}+V^{i(2)})+\langle p^i\rangle\\
P_{0(1)}+P_{i(1)}=\langle E\rangle(V_{0(2)}+V_{i(2)})-\langle p_i\rangle
\end{cases}
,\ \ \ \langle p_i\rangle\equiv\delta_{ij}\langle p^j\rangle
\end{equation}
in which the second formula is derived from a contraction between the first one and the metric $\eta_{\alpha\beta}$, with $P^{\alpha(1)}$ the instantly observed 4-momentum of $\mathtt{p_{(1)}}$ and $V^{\alpha(2)}$ the instantly observed 4-velocity of $\mathtt{p}_{(2)}$ that $V^{i(2)}=0$ iff $\mathtt{p}_{(2)}$ is at rest, and the index $i$ is running from $1$ to $3$. Multiply these two formulas above, it appears
\begin{equation}
P^{\alpha(1)}P_{\alpha(1)}=\langle E\rangle^2-\langle p\rangle^2=(m^{(1)})^2
\end{equation}
i.e. $E^2=m^2c^4+p^2c^2\ (c=1)$. Directly, one could figure out the relative velocity of an observed massive particle $\mathtt{p}_{(1)}$ from another massive particle $\mathtt{p}_{(2)}$ as the observer. All $\mathtt{p}_{(1)}$ with zero norm of relative 3-velocity $\beta\equiv\beta^{(2-1)}=0$ to $\mathtt{p}_{(2)}$ are in the rest frame of $\mathtt{p}_{(2)}$, or saying, \textit{at rest}. We could consider the mass of each $\mathtt{p}_{(1)}$ as its \textit{rest mass} for $\mathtt{p}_{(1)}$ is in the rest frame of itself. The mass of $\mathtt{p}_{(1)}$ is invariant whenever $\mathtt{p}_{(1)}$ receives a non-zero relative velocity $V^{\alpha(2-1)}$ (denoted below as $V^{\alpha(1)}$). To see this, we find the 4-velocity of $\mathtt{p}_{(2)}$ as the observer at rest that $P^{\alpha(2)}=m^{(2)}V^{\alpha(2)}$ and $V^{0(2)}=1$ then
\begin{equation}
\begin{split}
(d\tau)^2 &=(dx^{0(1)})^2-(dx^{i(1)})^2=(dx^{0(2)})^2-(dx^{i(2)})^2\\
&= (1-(\beta^{(1)})^2)(dx^{0(1)})^2=(dx^{0(2)})^2\\
&= (\dfrac{1}{(\beta^{(1)})^2}-1)^2(dx^{i(1)})^2\rightarrow0^+
\end{split}
\end{equation}
leading to
\begin{equation}
V^{0(1)}=\dfrac{1}{\sqrt{1-(\beta^{(1)})^2}}V^{0(2)}\equiv\gamma^{(1)}V^{0(2)}=\gamma^{(1)}
\end{equation}
where $\gamma\equiv\dfrac{1}{\sqrt{1-\beta^2}}$ and from $V^{i(1)}=V^{0(1)}\beta^{i(1)}$, it appears\footnote{When reproducing $V^{i(1)}$ by $V^{i(2)}$, one should omit an infinity as $V^{i(1)}=V^{0(1)}\beta^{i(1)}=\gamma^{(1)}\beta^{(1)}V^{i(2)}/\infty$. If $x^\alpha\in S_\alpha$ is continuous, then the infinity $\infty=\aleph_k\ (k\in\mathbb{Z}_+)$, else if $x^\alpha\in S_\alpha$ is discrete, then $\infty=\aleph_0=\vert\mathbb{N}\vert$. This case also happens in Eq.$(2.6)$, i.e. $(dx^{0(2)})^2-(dx^{i(2)})^2=(\dfrac{1}{(\beta^{(2)})^2}-1)^2(dx^{i(2)})^2=(dx^{i(2)})^2/\infty^2$.}
\begin{equation}
\begin{split}
P^{\alpha(1)}&=(m^{(1)}V^{0(1)}, m^{(1)}V^{i(1)})\\
&=(m^{(1)}\gamma^{(1)}, m^{(1)}\gamma^{(1)}\beta^{i(1)})
\end{split}
\end{equation}
exactly the correct form of 4-momentum stated in SR. An observed mass from 1+3 decomposition will often change to $\langle m\rangle=m\gamma$, but we care more for the invariant, authentic rest mass of each massive particle $\mathtt{p}$.

For massless particles, their masses are $m=0$ but they also evaluate along the proper time $\tau$ under the constraint Eq.$(2.3)$ in SR. Among them, if a massless particle has non-zero momentum $P^0>0$, then it must travel with invariant relative velocity $\vert V^0\vert\rightarrow+\infty$ or $\beta=1$, i.e. the speed of light in vacuum, thus it is instantaneous for $d\tau^2=0$ and cannot be referred to any specific coordinates $\{x^\alpha\}$. Else, if a massless particle has zero momentum $P^0=0$, then it may travel with arbitrary relative velocity $\vert V^0\vert>0$ or $\beta>0$, to be investigated in Section \ref{sec:4}.

\section{Geometric SR as quadratic space}

Now we turn to the geometry of SR along $\tau$, preparing for its quantization.

Find four generators $\gamma_\alpha$ over field $F=\mathbb{R},\ \mathbb{Z},\ \mathbb{R}\verb"\"\mathbb{R}_-$ or others to cover the domain of covariant CM-coordinate $\{x^\alpha\}=\{F\cdot\gamma_\alpha\}$. With an inner product $"\cdot":\ \gamma_\alpha\cdot\gamma_\beta=(\eta)_{\alpha\beta}\in\mathbb{R}$ being consistent with the metric $\eta_{\alpha\beta}$, vectors $\gamma_\alpha$ generate the real regular quadratic space $\mathbb{R}^{1,3}$ as its orthonormal basis in which the symmetric bilinear form is exactly the inner product $B(\gamma_\alpha, \gamma_\beta)=\gamma_\alpha\cdot\gamma_\beta=(\eta)_{\alpha\beta}$, and its quadratic form is $Q(\gamma_\alpha)\equiv\Vert\gamma_\alpha\Vert^2=B(\gamma_\alpha, \gamma_\alpha)=(\eta)_{\alpha\alpha}\in\{1, -1\}$ where $\Vert\gamma_\alpha\Vert$ are the norms of basis vectors $\gamma_\alpha$ of the inner product space of $\{x^\alpha\}=\mathbb{R}^{1,3}$,
\begin{equation}
Q(\gamma_0)=\Vert\gamma_0\Vert^2=1\ \ \ ,\ \ \ Q(\gamma_i)=\Vert\gamma_i\Vert^2=-1.
\end{equation}
In 3-vector description on an observed massive particle $\mathtt{p}$, we denote the unit vectors as $\overrightarrow{t}\equiv\gamma_0$, $\overrightarrow{x}^i\equiv\gamma_i$ for
\begin{equation}
\overrightarrow{\beta}^i\equiv\vert\dfrac{dx^i}{dx^0}\vert\overrightarrow{x}^i\ \ \ ,\ \ \ \overrightarrow{p}^i\equiv m\vert\dfrac{dx^i}{dx^0}\vert\overrightarrow{x}^i
\end{equation}
the vector components of its instant 3-velocity $\overrightarrow{\beta}\equiv\overrightarrow{\beta}^1+\overrightarrow{\beta}^2+\overrightarrow{\beta}^3$ and 3-momentum $\overrightarrow{p}\equiv\overrightarrow{p}^1+\overrightarrow{p}^2+\overrightarrow{p}^3$, as well as
\begin{equation}
\overrightarrow{V}^0\equiv\vert\dfrac{dx^0}{d\tau}\vert\overrightarrow{t}\ \ \ ,\ \ \ \overrightarrow{V}^i\equiv\vert\dfrac{dx^i}{d\tau}\vert\overrightarrow{x}^i
\end{equation}
the vector components of its instant 4-velocity $\overrightarrow{V}\equiv\overrightarrow{V}^0+\overrightarrow{V}^1+\overrightarrow{V}^2+\overrightarrow{V}^3$, and exact analogy for 4-momentum $\overrightarrow{P}$. Since every two different basis vectors $B(\gamma_\alpha, \gamma_\beta)=\gamma_\alpha\cdot\gamma_\beta=0\ (\alpha\neq\beta)$ are orthogonal, we have
\begin{equation}
\vert\overrightarrow{\beta}\vert=\beta\ \ \ \text{and}\ \ \ \vert\overrightarrow{p}\vert=p.
\end{equation}

Now we introduce a real Clifford algebra $\mathbb{R}^{1,3}\cong Cl^1(1,3)$ (isomorphic to $\mathbb{R}^{3,1}\cong Cl^1(3,1)$) of complex (Dirac) or real (Majorana), trivial matrix representations that $\gamma_\alpha^2=Q(\gamma_\alpha)\mathbf{1}$ and $\gamma_\alpha\gamma_\beta=-\gamma_\beta\gamma_\alpha\ (\alpha\neq\beta)$ with ordinary matrix product $\gamma_\alpha\gamma_\beta$ between $\gamma_\alpha$ and $\gamma_\beta$. Clifford products are then defined as
\begin{equation}
\begin{cases}
\gamma_\alpha\cdot\gamma_\beta=\gamma_{(\alpha}\gamma_{\beta)}=\dfrac{1}{2}(\gamma_\alpha\gamma_\beta+\gamma_\beta\gamma_\alpha)=\eta_{\alpha\beta}=\text{diag}(+,-,-,-)\\
\gamma_\alpha\times\gamma_\beta=\gamma_{[\alpha}\gamma_{\beta]}=\dfrac{1}{2}(\gamma_\alpha\gamma_\beta-\gamma_\beta\gamma_\alpha)=\gamma_{\alpha\beta}\in Cl^2(1,3)
\end{cases}
\end{equation}
where $\gamma_\alpha\cdot\gamma_\beta$ is the same as inner product. Dirac matrices of $\gamma_\alpha$ is constructed via Kronecker product $"\otimes"$ from Pauli matrices of a smaller Clifford algebra $Cl^1(3)\cong Cl^2(3)\cong spin(3)\cong su(2)$ that
\begin{equation}
\gamma_0=
\begin{pmatrix}
0 & 0 & 1 & 0\\
0 & 0 & 0 & 1\\
1 & 0 & 0 & 0\\
0 & 1 & 0 & 0
\end{pmatrix}
\ \gamma_1=
\begin{pmatrix}
0 & 0 & 0 & -1\\
0 & 0 & -1 & 0\\
0 & 1 & 0 & 0\\
1 & 0 & 0 & 0
\end{pmatrix}
\ \gamma_2=
\begin{pmatrix}
0 & 0 & 0 & i\\
0 & 0 & -i & 0\\
0 & -i & 0 & 0\\
i & 0 & 0 & 0
\end{pmatrix}
\ \gamma_3=
\begin{pmatrix}
0 & 0 & -1 & 0\\
0 & 0 & 0 & 1\\
1 & 0 & 0 & 0\\
0 & -1 & 0 & 0
\end{pmatrix}
\end{equation}
and generators of $Cl^2(1,3)$ are constructed via Eq.$(3.5)$,
\begin{equation}
\begin{split}
\gamma_{01}=
\begin{pmatrix}
0 & 1 & 0 & 0\\
1 & 0 & 0 & 0\\
0 & 0 & 0 & -1\\
0 & 0 & -1 & 0
\end{pmatrix}
\ \ \ \gamma_{02}=
\begin{pmatrix}
0 & -i & 0 & 0\\
i & 0 & 0 & 0\\
0 & 0 & 0 & i\\
0 & 0 & -i & 0
\end{pmatrix}
\ \ \ \gamma_{03}=
\begin{pmatrix}
1 & 0 & 0 & 0\\
0 & -1 & 0 & 0\\
0 & 0 & -1 & 0\\
0 & 0 & 0 & 1
\end{pmatrix}\\
\gamma_{12}=
\begin{pmatrix}
-i & 0 & 0 & 0\\
0 & i & 0 & 0\\
0 & 0 & -i & 0\\
0 & 0 & 0 & i
\end{pmatrix}
\ \ \ \gamma_{23}=
\begin{pmatrix}
0 & -i & 0 & 0\\
-i & 0 & 0 & 0\\
0 & 0 & 0 & -i\\
0 & 0 & -i & 0
\end{pmatrix}
\ \ \ \gamma_{31}=
\begin{pmatrix}
0 & -1 & 0 & 0\\
1 & 0 & 0 & 0\\
0 & 0 & 0 & -1\\
0 & 0 & 1 & 0
\end{pmatrix}
\end{split}
\end{equation}
where $\gamma_{03}$ and $\gamma_{12}$ are diagonal, to generate the Cartan subalgebra of $Cl^2(1,3)\cong spin(1,3)$ with Clifford products
\begin{equation}
\begin{split}
\gamma_{\alpha\beta}\cdot\gamma_{\gamma\delta} &= (\eta_{\alpha\delta}\eta_{\beta\gamma}-\eta_{\alpha\gamma}\eta_{\beta\delta})+\gamma_{\alpha\beta\gamma\delta}\\
\gamma_{\alpha\beta}\times\gamma_{\gamma\delta} &= \eta_{\alpha\delta}\gamma_{\beta\gamma}-\eta_{\alpha\gamma}\gamma_{\beta\delta}+\eta_{\beta\gamma}\gamma_{\alpha\delta}-\eta_{\beta\delta}\gamma_{\alpha\gamma}
\end{split}
\end{equation}
Noticing that all $\gamma$ generators are orthonormal only when $\Vert\gamma\Vert=1$, we should always rescale $\gamma\equiv\gamma/4$. In other descriptions, one may consider $J_{\alpha\beta}\equiv i\gamma_{\alpha\beta}/4$ as the orthonormal generator of Lorentz Lie algebra $spin(3,1)\cong spin(1,3)\cong su(2)\oplus su(2)$. If we are interested in de Sitter space \cite{73,74,75,76,77,78,79,80,81,82} of MacDowell-Mansouri gravity, we could find another generator $\gamma_4\in Cl^1(1,4)\verb"\"Cl^1(1,3)$ and $\gamma_4=\text{diag}(i,i,-i,-i)$ in matrix representation, which gives
\begin{equation}
\gamma_{04}=
\begin{pmatrix}
0 & 0 & -i & 0\\
0 & 0 & 0 & -i\\
i & 0 & 0 & 0\\
0 & i & 0 & 0
\end{pmatrix}
\ \gamma_{14}=
\begin{pmatrix}
0 & 0 & 0 & i\\
0 & 0 & i & 0\\
0 & i & 0 & 0\\
i & 0 & 0 & 0
\end{pmatrix}
\ \gamma_{24}=
\begin{pmatrix}
0 & 0 & 0 & 1\\
0 & 0 & -1 & 0\\
0 & 1 & 0 & 0\\
-1 & 0 & 0 & 0
\end{pmatrix}
\ \gamma_{34}=
\begin{pmatrix}
0 & 0 & i & 0\\
0 & 0 & 0 & -i\\
i & 0 & 0 & 0\\
0 & -i & 0 & 0
\end{pmatrix}
\end{equation}
together with $\gamma_{\alpha\beta}$, the generators of of rank-2 $spin(1,4)\cong Cl^2(1,4)\cong Cl^2(1,3)\oplus\{\gamma_4\}\times Cl^1(1,3)$. In some modern approaches \cite{22,91}, the four de Sitter generators are derived as $N_0=\gamma_{04}/4$ and $N_i=(\gamma_{i0}-\gamma_{i4})/4$, since the generator $\gamma_4$ is diagonal.

\section{Realizing SR in rank-4 spin algebras}
 \label{sec:4}
 
Here I want to present the realization of SR, starting with a discrete frame of $Cl^1(1,3)$, before the introduction of regularized 4-dimensional Cartan subalgebras.

Considering a massive particle $\mathtt{p}_{(1)}$ with its CM-coordinate $\{x^\alpha\},\ x^\alpha\in S_\alpha$ under the constraint Eq.$(2.3)$ in SR (now in SI units), we have
\begin{equation}
\begin{split}
(d\tau)^2 &= dx^0dx_0+\dfrac{dx^idx_i}{c^2} = (dx^0)^2-\sum_i\dfrac{(dx^i)^2}{c^2}\\
&= (1-\dfrac{v^2}{c^2})(dx^0)^2 = \dfrac{c^2-v^2}{c^2}(dx^0)^2\\
&= (\dfrac{1}{v^2}-\dfrac{1}{c^2})\sum_i(dx^i)^2=\dfrac{c^2-v^2}{v^2c^2}\sum_i(dx^i)^2
\end{split}
\end{equation}
for only its instant move at $d\tau\equiv\tau'-\tau=1,\ \tau\in\mathbb{N}$ with its 3-velocity $(v^i)^2\equiv(\beta^i)^2c^2\in\mathbb{N}$, or an observed 3-velocity from a $\mathtt{p}_{(2)}$ of $(d\tau)_{(1)}^2=(d\tau)_{(2-1)}^2$ that arranged universally,
\begin{equation}
(dx^{\alpha(2-1)})^2\equiv(dx^\alpha)^2\equiv(x^\alpha(\tau')-x^\alpha(\tau))^2\in\mathbb{N}.
\end{equation}
Thus from Eq.$(4.1)$, the difference on coordinates $\{x^\alpha\}$ are $(dx^0)^2=\dfrac{c^2}{c^2-v^2}(d\tau)^2$ and $\sum_i(dx^i)^2=\dfrac{v^2c^2}{c^2-v^2}(d\tau)^2$. Moreover, when the instant 3-velocity $\overrightarrow{v}$ of $\mathtt{p}$ is parallel to one of the orthonormal axes $\overrightarrow{x}^i$, there are $(dx^i)^2=\dfrac{v^2c^2}{c^2-v^2}(d\tau)^2$ for all $i=1,2,3$. To make these relations above be consistent with Eq.$(4.2)$, here we introduce necessarily the constant $k$ of $k^2\in\mathbb{Z}_+$ that
\begin{equation}
k^2(d\tau)^2=dx^0dx_0+\dfrac{dx^idx_i}{c^2}
\end{equation}
for correspondingly observed
\begin{equation}
\forall\ v^2\in\mathbb{N}\cap[0,c^2):\ 
\begin{cases}
(dx^0)^2=k^2\dfrac{c^2}{c^2-v^2}(d\tau)^2=k^2\dfrac{c^2}{c^2-v^2}\in\mathbb{Z}_+\\
(dx^i)^2=k^2\dfrac{v^2c^2}{c^2-v^2}(d\tau)^2=k^2\dfrac{v^2c^2}{c^2-v^2}\in\mathbb{N}
\end{cases}
.
\end{equation}

We know $c^2\in\mathbb{Z}_+>1$ for there exists $v\in(0,c)$. When $c^2$ is a prime, the constant $k$ should be at least
\begin{equation}
(k^2)_{min}=[1,\ 2,\ ...\ ,\ c^2-2,\ c^2-1]
\end{equation}
where the square bracket $[1,\ 2,\ ...\ ,\ c^2-2,\ c^2-1]$ stands for the least common multiple of positive integers $1,\ 2,\ ...\ ,\ c^2-2,\ \text{and}\ c^2-1$, while all selections of $k=h\cdot k_{min}\ (h^2\in\mathbb{Z}_+)$ are now technically correct, e.g. we could suppose an eligible $k$ as
\begin{equation}
k^2=\prod_{v^2=1}^{c^2-1}(c^2-v^2)\equiv\prod_{w=1}^{c^2-1}w=(c^2-1)!\ .
\end{equation}
When $c$ is not a prime, a minimal $k$ may be smaller. For each $v^2\in\mathbb{N}\cap(0,c^2)$ we define a  greatest common divisor of $(v^2,c^2)\equiv d_v$ and
\begin{equation}
\begin{cases}
v^2\equiv d_vm_v\\
c^2\equiv d_vn_v
\end{cases}
,\ \ \ (m_v,n_v)=1\ \ \ ,\ \ \ m_v,n_v\in\mathbb{Z}_+.
\end{equation}
From
\begin{equation}
\forall\ v^2\in\mathbb{N}\cap(0,c^2):\ k^2\dfrac{c^2}{c^2-v^2}\in\mathbb{Z}_+\ \ \ \Rightarrow\ \ \ c^2k^2\ \vert\ c^2-v^2\ \ \ \Rightarrow\ \ \ d_vn_vk^2\ \vert\ d(n_v-m_v)
\end{equation}
we obtain $\forall\ v:\ n_vk^2\ \vert\ n_v-m_v$ and $(m_v,n_v)=1$ hence $k^2\ \vert\ n_v-m_v$ that
\begin{equation}
(k^2)_{min}=[n_1-m_1,\ n_2-m_2,\ ...\ ,\ n_{c^2-2}-m_{c^2-2},\ n_{c^2-1}-m_{c^2-1}]
\end{equation}
and obviously, $k^2=(c^2-1)!$ is as well an eligible selection. Since if $v^2\in\mathbb{Q}_+$ then $k^2=\aleph_0$ is not acceptable, above in our discrete arrangement, $\tau\in\mathbb{N}$ and for every instant movement $d\tau=1$ of a particle, $\forall\ \alpha: (dx^\alpha)^2\in\mathbb{N}$ thus $(V^\alpha)^2\in\mathbb{N}$, we set up the relative $(v^i)^2\in\mathbb{N}$, $v^2\in\mathbb{N}\cap[0,c^2]$, $c\in\mathbb{Z}_+>1$, and $k^2\in\mathbb{Z}_+$. However, in current description, the instantly observed $dx^\alpha$, $V^\alpha$, $v^i$, $v$, $c$, and $k$ may not necessarily be integers. They live in the set $\pm\sqrt{\mathbb{N}}$ which is discrete as well, with properties in Eq.$(3.4)$ derived from their orthonormal vector-like definition. The constant $k$ is important to the embedded, discrete base of 4-dimensional Cartan subalgebras.

Another property is found when regarding $(dx^0)^2$ or $(dx^i)^2$ as functions of correspondingly the instant 3-velocity of $\mathtt{p}$ as
\begin{equation}
\begin{cases}
(dx^0)^2(v^2)\varpropto f_0(\beta^2)\equiv\dfrac{1}{1-\beta^2}\\
(dx^i)^2(v^2)\varpropto f_i(\beta^2)\equiv\dfrac{\beta^2}{1-\beta^2}
\end{cases}
\end{equation}
which may be relevant to matrices $\Lambda_\beta^\alpha$ of Lorentz transform $x'^\alpha=\Lambda_\beta^\alpha x^\beta$ including spin and boost, e.g.
\begin{equation}
\Lambda_\beta^\alpha[\beta^1]=\begin{pmatrix}
\gamma & -\beta\gamma & 0 & 0\\
-\beta\gamma & \gamma & 0 & 0\\
0 & 0 & 1 & 0\\
0 & 0 & 0 & 1
\end{pmatrix}
\end{equation}
indicating that the functions $(dx^\alpha)^2$ of $v$ or $v^2$ are not linear, but hyperbolic.

\subsection{4-Dimensional Cartan base}

Consider a 4-dimensional regular quadratic space $S^4$ over domain $S$ with metric $\delta_{\alpha\beta}=\text{diag}(+,+,+,+)$ and its orthonormal basis vectors $c_\alpha$ of $c_\alpha^2=1$ and $c_\alpha c_\beta=0\ (\alpha\neq\beta)$, in which a set of real numbers as the \textit{locations} $\{s^\alpha\}=\{s_\alpha\}$ for $\{ds_\alpha\}=\{S\cdot c_\alpha\}$ of points $\mathtt{P}$ are either continuous or discrete, determined by the cardinal number of domain $S$, i.e. for a location of some specified $\mathtt{P}$, $\forall\ \alpha:(ds^\alpha)^2(\mathtt{P})\in\mathbb{R}\verb"\"\mathbb{R}_-$ or $\mathbb{N}$. Firstly we embed the spacetime of $Cl^1(1,3)$ onto the base $F^4\cong Cl^1(4)$ for discovering structural properties in SR.

On a continuous level, the tangent space $Cl^1(1,3)\cong\mathbb{R}^{1,3}$ is embedded on $Cl^1(4)\cong\mathbb{R}^{4}$, while the connection of Clifford basis vectors $c_\alpha$ and locations $\{ds^\alpha\}$ of points $\mathtt{P}$ appear as a Maurer-Cartan 1-form over the tangent space
\begin{equation}
\underline{c}=d\underline{s}^\alpha\delta_\alpha^{\ \beta}(s)\ c_\beta=d\underline{s}^\alpha c_\alpha
\end{equation}
with $\delta_\alpha^{\ \beta}(s)=\text{diag}(+,+,+,+)$ for all considered locations $s\equiv s^\alpha$ in $\mathbb{R}^4$. Over the embedded manifold $Cl^1(1,3)\cong\mathbb{R}^{1,3}$ of the spacetime, the connection of Clifford basis vectors $\gamma_\alpha$ and real CM-coordinates $\{dx^\alpha\}$ of massive particles $\mathtt{p}$ is described by the frame of $\underline{e}\in C\underline{l}^1(1,3)$ appearing as a Maurer-Cartan 1-form over the tangent space
\begin{equation}
\underline{e}=d\underline{x}^\alpha\eta_\alpha^{\ \beta}(x)\ \gamma_\beta=d\underline{x}^\alpha\gamma_\alpha
\end{equation}
for at every points in the flat Minkowski spacetime, $\gamma_\alpha=\eta_{\alpha\beta}\gamma^\beta=\eta_{\alpha\beta}\eta^{\beta\gamma}\gamma_\gamma=\eta_\alpha^{\ \gamma}\gamma_\gamma$ that $\eta_\alpha^{\ \beta}=\delta_\alpha^{\ \beta}=\text{diag}(+,+,+,+)$, we can find the geometric relations between connections $\underline{c}$ and $\underline{e}$ locally as
\begin{equation}
\underline{c}=d\underline{s}^\alpha c_\alpha=d\underline{s}^\alpha\bar{t}_\alpha^{\ \beta}(s)\ \gamma_\beta\ \ \ \ \ \ \underline{e}=d\underline{x}^\alpha\gamma_\alpha=d\underline{x}^\alpha t_\alpha^{\ \beta}(x)\ c_\beta
\end{equation}
with
\begin{equation}
ds^\beta=dx^\alpha t_\alpha^{\ \beta}(x)\ \ \ \ \ \ \gamma_\alpha=c_\beta\ t_\alpha^{\ \beta}(x)
\end{equation}
an invertible matrix $t_\alpha^{\ \beta}(x)$ satisfying $t_\alpha^{\ \beta}\bar{t}_\beta^{\ \gamma}=\delta_\alpha^{\ \gamma}$, which is able to transform between differentiated $\{s^\alpha\}$ and $\{x^\alpha\}$ locally. In most cases, the embedder $t_\alpha^{\ \beta}(x)$ is globally constant and diagonal.

In the same way, some conformal properties of curved spacetime $dx^\mu\equiv dx^\alpha e_\alpha^{\ \mu}$ with an invertible matrix $e_\mu^{\ \alpha}(x)$ satisfying $e_\mu^{\ \alpha}e_\alpha^{\ \nu}=\delta_\mu^\nu$ and $e_\mu^{\ \alpha}e_\beta^{\ \mu}=\delta_\beta^\alpha$ of the connection 1-form
\begin{equation}
\underline{e}=d\underline{x}^\alpha\gamma_\alpha=d\underline{x}^\mu e_\mu^{\ \alpha}(x)\ \gamma_\alpha
\end{equation}
can also be derived from the frame $\underline{e}$ of $\mathbb{R}^{1,3}$ in SR, as tetrad $\overrightarrow{e}_\alpha(x)=e_\alpha^{\ \mu}(x)\overrightarrow{\partial}_\mu$ and cotetrad $\underline{e}^\alpha(x)=d\underline{x}^\mu e_\mu^{\ \alpha}(x)$ satisfying $\overrightarrow{e}_\alpha\underline{e}^\beta=\delta_\alpha^\beta$ are constructed to raise
\begin{equation}
(\overrightarrow{e}_\alpha, \overrightarrow{e}_\beta)\equiv e_\alpha^{\ \mu}g_{\mu\nu}(x)e_\beta^{\ \nu}=\eta_{\alpha\beta}(x)=\eta_{\alpha\beta}=\text{diag}(+,-,-,-)
\end{equation}
that locally $\overrightarrow{e}_\alpha(x)\simeq\gamma_\alpha$ where $g_{\mu\nu}(x)$ is the metric of curved spacetime, the conformal manifold of $\mathbb{R}^{1,3}$. In torsionless de Sitter space \cite{73,74,75,76,77,78,79,80,81,82}, we have $e_\mu^{\ \alpha}=\text{diag}(1,e^{t/l},e^{t/l},e^{t/l})$ with $\alpha=1/l$ the expansion parameter. That the 4-dimensional line element $(d\tau_S)^2=(dx^0)^2-\sum_i(e^{2t/l})(dx^i)^2/c^2$ of de Sitter space can be derived from five orthogonal generators $\gamma_\alpha$ and $\gamma_4$ is for the spin representation of generator $\gamma_4$ above Eq.$(3.9)$ is diagonal.

On a discrete level, noticing that $v^2\in\mathbb{Q}_+\Rightarrow k^2 = \aleph_0$ is unacceptable, here I want to construct a discrete frame $\mathbb{N}^4$ generated by four orthonormal vectors $c_\alpha$, in which each element of the space $\forall\ \mathtt{e}\in\mathbb{N}^4$ corresponds to four elements $n^0, n^1, n^2$, and $n^3$, each $n^\alpha\in\mathbb{N}$ on $\{\mathbb{N}\cdot c^\alpha\}$, has a unique \textit{position} $\{n^\alpha(\mathtt{e})\}$ of $n^\alpha(\mathtt{e})\in\mathbb{N}$ in $\mathbb{N}^4$. To realize the discrete arrangements at each $\tau$ and leave our considered particles $\mathtt{p}$ in the space $\mathbb{R}^{1,3}$, a (tangent) discrete frame $e$ from another four orthogonal axes of $\{\mathbb{N}\cdot\gamma_\alpha\}$ should be quantized as massless, zero-momentum, split particles $e_{a(i)}$ distributing on $\{\mathbb{N}\cdot c_\alpha\}$ with $a=\text{O}, \text{I}, \text{II}, \text{III}\leftrightarrow\alpha=0, 1, 2, 3$. Each\footnote{To arrange the labels $i\in\mathbb{N}$ of quantized frame embedded on $\mathbb{N}^4$, we may select $i\equiv0\ (\text{mod}\ 4)$ for $e_{\text{O}(i)}$, $i\equiv1\ (\text{mod}\ 4)$ for $e_{\text{I}(i)}$, $i\equiv2\ (\text{mod}\ 4)$ for $e_{\text{II}(i)}$, and $i\equiv3\ (\text{mod}\ 4)$ for $e_{\text{III}(i)}$. The well-order on $\mathbb{N}^4$ and $\mathbb{N}^{1,3}$ are natural, i.e. $j>i\in\mathbb{N}\Leftrightarrow n^\alpha(e_{a(j)})>n^\alpha(e_{a(i)})$.} of $e_{\text{O}(h)}\in\{\mathbb{N}\cdot c_0\}$, $e_{\text{I}(i)}\in\{\mathbb{N}\cdot c_1\}$, $e_{\text{II}(j)}\in\{\mathbb{N}\cdot c_2\}$, and $e_{\text{III}(k)}\in\{\mathbb{N}\cdot c_3\}$, $\forall\ h, i, j, k\in\mathbb{N}$ has a unique position in $\mathbb{N}^4$, i.e. $\{n^\alpha(e_{a(i)})\}$ of $n^\alpha(e_{a(i)})\in\mathbb{N}$.

After an eligible observation has been selected as in all the discussions below, each of the massive particles $\mathtt{p}_{(q)}$, $\forall\ q\in\mathbb{N}$ addressed in $\mathbb{R}^{1,3}$ with its unique CM-coordinate $\{x^{\alpha(q)}\}\equiv\{x^\alpha(\mathtt{p}_{(q)})\}$ in SR corresponds to a unique position of $n^\alpha(\mathtt{p}_{(q)})=n^\alpha(e_{a(i)})$ at each $\tau$ while $\mathtt{p}_{(q)}$ is \textit{related} to $e_{a(i)}$. Denote
\begin{equation}
n^\alpha(e_{a(j-i)})\equiv n^\alpha(e_{a(j)})-n^\alpha(e_{a(i)})\in\mathbb{Z}
\end{equation}
as the difference between every two split particles $e_{a(j)}$ and $e_{a(i)}$ of the quantized frame. Considering the natural well-order on every $\mathbb{N}\subset\mathbb{N}^4$, the difference $n^\alpha(e_{a(i+4\ -\ i)})$ can be defined between two adjacent $e_{a(i+4)}$ and $e_{a(i)}$ iff $\nexists\ j\in\mathbb{N}: n^\alpha(e_{a(i+4)})>n^\alpha(e_{a(j)})>n^\alpha(e_{a(i)})$. If we assume that in most cases on each $\alpha$ and $a$, the distribution of $e_{a(i)}$ on $c_\alpha$ is even, i.e. $\forall\ i\in\mathbb{N}$,
\begin{equation}
n^\alpha(e_{a(i+4\ -\ i)})=\text{const}\in\mathbb{N}.
\end{equation}

Find $d^\alpha(e_{a(i)})$ under our arrangement above, of
\begin{equation}
d^\alpha(e_{a(i)})\equiv[\dfrac{i}{4}]\in\mathbb{N}
\end{equation}
where the bracket $[\ \ ]$ stands for the Gaussian Internal Function and
\begin{equation}
d^\alpha(e_{a(j-i)})\equiv d^\alpha(e_{a(j)})-d^\alpha(e_{a(i)})=\dfrac{j-i}{4}\in\mathbb{Z}
\end{equation}
when $j\equiv i$ (mod 4),
\begin{equation}
(dx^{\alpha(q)})^2(\tau)=d^\alpha(e_{a(i)})=[\dfrac{i}{4}]
\end{equation}
appears the instant difference $dx^{\alpha(q)}$ on the observed CM-coordinate $\{x^{\alpha(q)}\}$ at each $d\tau\equiv\tau'-\tau=1$ while $\mathtt{p}_{(q)}$ is related to $e_{a(i)}$ at $\tau\in\mathbb{N}$. Hence
\begin{equation}
(U^{\alpha(q)})^2(\tau)=\dfrac{(dx^{\alpha(q)})^2(\tau)}{(d\tau)^2}=d^\alpha(e_{a(i)})\in\mathbb{N}
\end{equation}
appears its instant, pseudo- 4-velocity $U^{\alpha(q)}(\tau)$, raising
\begin{equation}
v^{i(q)}(\tau)\equiv\dfrac{dx^{i(q)}(\tau)}{dx^{0(q)}(\tau)}=\dfrac{U^{i(q)}}{U^{0(q)}}
\end{equation}
where $i=1,2,3$ is an index. If we denote $l\equiv i$ (mod 4), $h\equiv 0$ (mod 4), $l, h\in\mathbb{N}$ as the labels of discrete frame $e_{i(l)}$ and $e_{\text{O}(h)}$ on $c_i$ and $c_0$, then it appears
\begin{equation}
(v^{i(q)})^2(\tau)=\dfrac{(U^{i(q)})^2}{(U^{0(q)})^2}=\dfrac{d^i(e_{i(l)})}{d^0(e_{\text{O}(h)})}=[\dfrac{l}{4}]\cdot\dfrac{4}{h}\in\mathbb{N}
\end{equation}
its instant 3-velocity $v^{i(q)}(\tau)$, as well as
\begin{equation}
p^{i(q)}(\tau)=m^{(q)}(\tau)\cdot v^{i(q)}(\tau)
\end{equation}
its 3-momentum. Instantly, we derive
\begin{equation}
(B^{\alpha(q)})^2(\tau)\equiv\dfrac{d^\alpha(e_{a(i'-i)})}{(d\tau)^2}=\dfrac{i'-i}{4}
\end{equation}
of its pseudo- 4-acceleration at $\tau\in\mathbb{N}$ iff $n^\alpha(\mathtt{p}_{(q)})(\tau')=n^\alpha(e_{a(i')})$, $n^\alpha(\mathtt{p}_{(q)})(\tau)=n^\alpha(e_{a(i)})$, and $\tau'-\tau=1$. Here $i'$ and $i$ are labels of $i'\equiv i\equiv\alpha$, $\alpha=0,1,2,3$. Hence, it appears
\begin{equation}
(b^{i(q)})^2(\tau)=\dfrac{(B^{i(q)})^2(\tau)}{(B^{0(q)})^2(\tau)}=\dfrac{d^i(e_{i(l'-l)})}{d^0(e_{\text{O}(h'-h)})}=\dfrac{l'-l}{h'-h}
\end{equation}
of its instant 3-acceleration $b^{i(q)}(\tau)$, iff $n^i(\mathtt{p}_{(q)})(\tau')=n^i(e_{i(l')})$, $n^i(\mathtt{p}_{(q)})(\tau)=n^i(e_{i(l)})$, $n^0(\mathtt{p}_{(q)})(\tau')=n^0(e_{\text{O}(h')})$, $n^0(\mathtt{p}_{(q)})(\tau)=n^0(e_{\text{O}(h')})$, and $\tau'-\tau=1$. Find
\begin{equation}
(ds^{\alpha(q)})^2(\tau)\equiv n^\alpha(e_{a(i)})\in\mathbb{N}
\end{equation}
at each $d\tau=\tau'-\tau=1$ while $\mathtt{p}_{(q)}$ is related to $e_{a(i)}$ at $\tau\in\mathbb{N}$, to obtain
\begin{equation}
(V^{\alpha(q)})^2(\tau)=\dfrac{(ds^{\alpha(q)})^2(\tau)}{(d\tau)^2}=n^\alpha(e_{a(i)})
\end{equation}
its instant, authentic 4-velocity $V^{\alpha(q)}(\tau)$, and
\begin{equation}
(a^{\alpha(q)})^2(\tau)=\dfrac{n^\alpha(e_{a(i'-i)})}{(d\tau)^2}=n^\alpha(e_{a(i'-i)})\in\mathbb{N}
\end{equation}
its authentic 4-acceleration at each $\tau\in\mathbb{N}$, with $n^\alpha(\mathtt{p}_{(q)})(\tau')=n^\alpha(e_{a(i')})$. From $V^i=V^0\beta^i\Leftrightarrow\dfrac{V^i}{V^0}=\dfrac{v^i}{c}$,
\begin{equation}
\dfrac{n^i(e_{i(l)})}{n^0(e_{\text{O}(h)})}=\dfrac{d^i(e_{i(l)})}{d^0(e_{\text{O}(h)})\cdot c^2}
\end{equation}
in most cases, under an even distribution of discrete frame related to $\mathtt{p}_{(q)}$ at $\tau\in\mathbb{N}$,
\begin{equation}
n^\alpha(e_{a(j-i)})=d^\alpha(e_{a(j-i)})\cdot n^\alpha(e_{a(i+4\ -\ i)})\ \ \ ,\ \ \ \forall\ j>i\in\mathbb{N},\ j\equiv i\ (\text{mod\ 4})
\end{equation}
it appears the continuous approximation of
\begin{equation}
(t_\alpha^{\ \alpha})^2(x)=n^\alpha(e_{a(i+4\ -\ i)})=\text{const}\in\mathbb{N}\ \ \ \text{(no\ sum)}
\end{equation}
and
\begin{equation}
\begin{cases}
(ds^\alpha)^2=n^\alpha(e_{a(i+4\ -\ i)})(dx^\alpha)^2\\
(V^\alpha)^2=n^\alpha(e_{a(i+4\ -\ i)})(U^\alpha)^2
\end{cases}
\end{equation}
that
\begin{equation}
\dfrac{n^i(e_{i(l+4\ -\ l)})}{n^0(e_{\text{O}(h+4\ -\ h)})}=\dfrac{1}{c^2}
\end{equation}
and
\begin{equation}
\dfrac{(t_i^{\ i})^2(x)}{(t_0^{\ 0})^2(x)}=\dfrac{1}{c^2}\ \ \ \text{(no\ sum)}
\end{equation}
leading to
\begin{equation}
\begin{split}
k^2(d\tau)^2 &= dx^0dx_0+\dfrac{dx^idx_i}{c^2}\\
&= (dx^0)^2-\sum_i\dfrac{(dx^i)^2}{c^2}\\
&= \dfrac{ds^0ds_0}{n^0(e_{\text{O}(h+4\ -\ h)})}-\dfrac{ds^ids_i}{n^i(e_{i(l+4\ -\ l)})\cdot c^2}\\
&= \dfrac{(ds^0)^2}{n^0(e_{\text{O}(h+4\ -\ h)})}-\dfrac{(ds^i)^2}{n^i(e_{i(l+4\ -\ l)})\cdot c^2}
\end{split}
\end{equation}
in which $c^2k^2\in\mathbb{N}$. For a valid selection, $c^2k^2=(c^2)!$. Continuing on $\mathtt{p}_{(q)}$, instantly
\begin{equation}
(dx^0)^2=k^2(d\tau)^2=k^2\ \ \ \Rightarrow\ \ \ d^0(e_{\text{O}(h)})=k^2=(c^2-1)!
\end{equation}
we have $n^0(e_{\text{O}(h)})=(c^2)!$, when $\mathtt{p}_{(q)}$ is in its rest frame that $V^{i(q)}=0$ and $n^i(e_{i(l+4\ -\ l)})=1$.

It is a truth that the associated structures in $\mathbb{N}^4$ as the position $\{n^\alpha\}$, 4-velocity $V^\alpha$, and 4-momentum $P^\alpha\equiv mV^\alpha$ are fixed, but the distributions of (tangent) frame are flexible. Consider a constant
\begin{equation}
\bar{c}^2=\dfrac{\bar{n}^0(e_{\text{O}(h+4\ -\ h)})}{\bar{n}^i(e_{i(l+4\ -\ l)})}
\end{equation}
of another even distribution of the frame,
\begin{equation}
\begin{cases}
(ds^i)^2 = \bar{n}^i(e_{i(l+4\ -\ l)})(d\bar{x}^i)^2\\
(ds^0)^2 = \bar{n}^0(e_{\text{O}(h+4\ -\ h)})(d\bar{x}^0)^2
\end{cases}
\end{equation}
thus a massive particle $\mathtt{p}_{(q)}$ at $\forall\ \tau\in\mathbb{N}$ with corresponding $\bar{v}^i\equiv\dfrac{d\bar{x}^i}{d\bar{x}^0}$ should satisfy
\begin{equation}
\dfrac{ds^i}{ds^0}=\dfrac{V^i}{V^0}=\dfrac{v^i}{c}=\dfrac{\bar{v}^i}{\bar{c}}
\end{equation}
that
\begin{equation}
\bar{v}^i(\tau)=\dfrac{\bar{c}}{c}v^i(\tau)
\end{equation}
under its conserved 4-velocity of $V^\alpha(\tau)$, leaving $(\bar{v}^i)^2\in\mathbb{N}\cap[0,\bar{c}^2)$ for all $\mathtt{p}_{(q)}$ under the distribution where Eqs.$(4.41),(4.42),(4.43)$ are eligible for all observed $\bar{v}^{i(r-q)}$ and $v^{i(r-q)}$.

Generally, an observer-dependent
\begin{equation}
n^{\alpha(r-q)}(e_{a(i)})=\sideset{}{_{\lambda\equiv\alpha\ \text{(mod\ 4)}}}\sum_{\lambda=0}^{i-4}n^{\alpha(r-q)}(e_{a(\lambda+4\ -\ \lambda)})
\end{equation}
is determined in an arbitrary distribution, while
\begin{equation}
n^{\alpha(r-q)}(e_{a(j-i)})=\sideset{}{_{\lambda\equiv\alpha\ \text{(mod\ 4)}}}\sum_{\lambda=i}^{j-4}n^{\alpha(r-q)}(e_{a(\lambda+4\ -\ \lambda)})
\end{equation}
with $j>i\in\mathbb{N}$, $j\equiv i$ (mod 4), are eligible for computing the constant $\bar{c}$ locally.

Applying the discrete frame only in tangent space, i.e. instantly iff $d\tau=\tau'-\tau=1$, we have the universally denoted $(ds^\alpha)^2$, $(dx^\alpha)^2$, $(V^\alpha)^2$, $(U^\alpha)^2$, $(v^i)^2$ and $c^2$, $k^2$ in $\mathbb{N}$, hence the quadratic space $F_\geq\oplus F^3\subset F^{1,3}$ generated by $\gamma_\alpha$ is valued in
\begin{equation}
F=\{\sum_{\lambda=1}^{\aleph_0}\pm\sqrt{a_\lambda}\}\ \ \ ,\ \ \ a_\lambda\in\mathbb{N}
\end{equation}
that $S_i\subset F$ and
\begin{equation}
S_0\subset F_\geq\equiv\{\sum_{\lambda=1}^{\aleph_0}\sqrt{a_\lambda}\}\neq F\verb"\" F_-\ \ \ ,\ \ \ a_\lambda\in\mathbb{N}
\end{equation}
of the cardinal
\begin{equation}
\vert F\vert=\vert F_\geq\vert=\aleph_0=\vert\mathbb{N}\vert
\end{equation}
which declares a distance between two different massive particles $\mathtt{p}_{(1)}$ and $\mathtt{p}_{(2)}$ at each $\tau$,
\begin{equation}
x^{\alpha(2-1)}(\tau)=\sum_{\lambda=\tau}^{\tau''}\pm\sqrt{\vert d^\alpha(e_{a(i)})(\lambda)\vert}=\dfrac{1}{2}\sum_{\lambda=\tau}^{\tau''}\pm\sqrt{\vert i(\lambda)\vert}
\end{equation}
where $x^{\alpha(2)}(\tau)=x^{\alpha(1)}(\tau'')$ and $\mathtt{p}_{(1)}$ is related to $e_{a(i)}(\lambda)$ at each $\lambda\in\mathbb{N}$. The distances between CM-coordinate of considered massive particles are alive only within the evolution of \textit{referring} (as above Eq.$(4.2)$), well-ordered proper time $\tau\in\mathbb{N}$.

Our space in SR might be a subset of continuous $Cl^1(1,3)\cong\mathbb{R}^{1,3}$, but our history lives in a smaller set $F_\geq\oplus F^3$ of
\begin{equation}
\vert F\vert=\vert F_\geq\vert<\vert\mathbb{R}\vert.
\end{equation}

\subsection{Orthogonal Regulation}

The quantized tangent frame $e_{a(i)}$ on $c_\alpha$ of $i\equiv a$ (mod 4) related only to every $\mathtt{p}$ are massless, zero-momentum particles with positions $\{n^\alpha(e_{a(i)})\}$ of $n^\alpha(e_{a(i)})\in\mathbb{N}$. Its (wave) function is anti-symmetric, since $\forall\ j\neq i:$
\begin{equation}
\Psi(e_{a(j-i)})\equiv n^\alpha(e_{a(j-i)})=-n^\alpha(e_{a(i-j)})=-\Psi(e_{a(i-j)}).
\end{equation}
The quantized domain $S_4\subset\pm\sqrt{\mathbb{N}}$ (or $\mathbb{R}$) of particle masses generated by $\gamma_4$ embedded on $c_4$ of $Q(c_4)=\Vert c_4\Vert^2=1$ in a dimension orthogonal to $\{\gamma_\alpha\}$, as $\{\gamma_4\}=Cl^1(1,4)\verb"\"Cl^1(1,3)$ in MacDowell-Mansouri gravity, produces massless, zero-momentum particles $\phi_{\text{IV}(r)}$, $r\in\mathbb{N}$ with positions $\{n^4(\phi_{\text{IV}(r)})\}$ of $n^4(\phi_{\text{IV}(r)})\in\mathbb{N}$ as the exact analogy of $e_{a(i)}$. Its (wave) function is anti-symmetric, since $\forall\ t\neq r:$
\begin{equation}
\Psi(\phi_{\text{IV}(t-r)})\equiv n^4(\phi_{\text{IV}(t-r)})=-n^4(\phi_{\text{IV}(r-t)})=-\Psi(\phi_{\text{IV}(r-t)}).
\end{equation}

A direct product between $\{\gamma_\alpha\}$ and $\{\gamma_4\}$ introduces the space $\{\gamma_{\alpha4}\}$ generated by $\gamma_{\alpha4}=\gamma_\alpha\times\gamma_4$ with a Clifford product in Eq.$(3.5)$, in which $\gamma_{\alpha4}\equiv T_A$ be considered as well as generators of $spin(1,4)$ if $Q(\gamma_4)=\Vert\gamma_4\Vert^2=-1$ or $spin(2,3)$ else if $Q(\gamma_4)=\Vert\gamma_4\Vert^2=1$. The quantized particles $e_a\phi_{\text{IV}(s)}$ as combined $e_{a(i)}$ and $\phi_{\text{IV}(r)}$ have symmetric (wave) functions, since $\forall\ \pi\neq s:$
\begin{equation}
\begin{split}
\Psi(e_a\phi_{\text{IV}(\pi-s)}) &\equiv n^\alpha(e_{a(i(\pi)-i(s))})\cdot n^4(\phi_{\text{IV}(r(\pi)-r(s))})\\
&= (-1)^2\cdot n^\alpha(e_{a(i(s)-i(\pi))})\cdot n^4(\phi_{\text{IV}(r(s)-r(\pi))})\\
&= \Psi(e_a\phi_{\text{IV}(s-\pi)}).
\end{split}
\end{equation}
Therefore, the particles $e_a\phi_{\text{IV}(s)}$ are bosons with $(s)$ labelling different $e_a\phi_\text{IV}$ in a discrete frame. The same as other untwisted \cite{86,87,88,89,90,93,94,95} massive particles $\mathtt{p}$, each $e_a\phi_{\text{IV}(s)}$ is \textit{related} to some split $e_a\phi_\text{IV}$ of $e_{a(i)}$ and $\phi_{\text{IV}(r)}$ at each $\tau\in\mathbb{N}$, receiving its unique 3-velocity, 4-velocity, and its unique rest mass of $(m^2)(e_a\phi_{\text{IV}(s)})\equiv x^4(e_a\phi_{\text{IV}(s)})=x^4(\phi_{\text{IV}(r)})$, i.e. being addressed on the tangent spaces in SR.

When considering de Sitter space of $spin(1,4)$, the particles $e_a\phi_4$ are called \textit{frame-Higgs} in MacDowell-Mansouri gravity. Together with $\underline{\omega}\sim C\underline{l}^2(1,3)$, frame-Higgs from $\underline{e}\phi\sim\{\gamma_4\}\times C\underline{l}^1(1,3)$ raise the gauged Lie algebra of $Cl^2(1,4)\cong spin(1,4)$ in which the connection 1-form is defined smoothly as $\underline{A}=\underline{\omega}+\underline{e}\phi$ to meet $\underline{\underline{T}}\phi=\underline{d}\underline{e}+\underline{\omega}\times\underline{e}\phi=0$ for pure de Sitter space.

In other cases, if there exists more than one generations of split $e_{a(i)}$ and $\phi_{\text{IV}(r)}$ related to a massive particle $\mathtt{p}$, then
\begin{equation}
\forall\ i'\neq i:\ n^\alpha(e_{a'(i')})=n^\alpha(e_{a(i)})\ \ \ \Leftrightarrow\ \ \ x^\alpha(e_{a'(i')})=x^\alpha(e_{a(i)})
\end{equation}
and
\begin{equation}
\forall\ r'\neq r:\ n^4(\phi_{\text{IV}'(r')})=n^4(\phi_{\text{IV}(r)})\ \ \ \Leftrightarrow\ \ \ x^4(\phi_{\text{IV}'(r')})=x^4(\phi_{\text{IV}(r)}).
\end{equation}
Noticing that the anti-symmetric (wave) functions of split $e_{a(i)}$ and $\phi_{\text{IV}(r)}$ are restricted by \textit{Pauli exclusion principle}, an equation of $n^\alpha(e_{a'(i')})=n^\alpha(e_{a(i)})$ on $c_\alpha$ leads to $i'=i$ if $a'=a$, and $n^4(\phi_{\text{IV}'(r')})=n^4(\phi_{\text{IV}(r)})$ leads to $r'=r$ if $\text{IV}'=\text{IV}$. Hence, the case of Eqs.$(4.53)$ and $(4.54)$ could happen only when more than one generations of split $e_{a'(i')}$ and $e_{a(i)}$, $\forall\ a'\neq a$ are defined separately from different generators $\{\gamma_\alpha'\}$ and $\{\gamma_\alpha\}$ of different (but isomorphic) gauged algebras $\gamma_{\alpha'}\in Cl^1(1,3)'$ and $\gamma_\alpha\in Cl^1(1,3)$ as subalgebras $Cl^1(1,3)'=L_g\cdot Cl^1(1,3)$ of a larger one, $g$, and so do $\phi_{\text{IV}'(r')}$ and $\phi_{\text{IV}(r)}$. Other conditions in these cases are in need of extra restrictions.

For each generation of $e_{a(i)}\sim\gamma_\alpha\in Cl^1(1,3)$ and $\phi_{\text{IV}(r)}\sim\gamma_4\in Cl^1(1,4)\verb"\"Cl^1(1,3)$, each generator $\gamma_\alpha$ or $\gamma_4$ is embedded only on one dimension of $c_\alpha$, while every two orthogonal generators of $\gamma_\alpha\cdot\gamma_\beta=0$ and $\gamma_\alpha\cdot\gamma_4=0$ cannot be embedded on the same dimension of $c_\alpha$ that we can directly arrange $a=\text{O},\text{I},\text{II},\text{III}\leftrightarrow\alpha=0,1,2,3$ and indices of $\text{IV}\leftrightarrow4$. The condition holds also for more than one generations, i.e. the \textit{orthogonal regulation} reads

\color{white}\ \color{black}

\textit{Every two orthogonal generators $\gamma_\alpha$ and $\gamma_\beta$ of $\gamma_\alpha\cdot\gamma_\beta=0$ corresponding to split particles in a gauged algebra $g$ cannot be embedded on the same dimension of the Cartan algebra of $g$.}

\color{white}\ \color{black}

\noindent Under the orthogonal regulation, we could specify our arrangements in the embeddings which involve more than one generations of split frame-Higgs. Besides, a massless particle $\mathtt{p}_{(q)}$ always corresponds to a closed region $[x^{\alpha(q)}_{in.}(\tau),x^{\alpha(q)}_{fi.}(\tau)]$ in 1-dimensional subspace of $\{F_\geq\cdot\gamma_0\}$ or $\{F\cdot\gamma_i\}$ in the spacetime at each $\tau\in\mathbb{N}$ but not related to any split frame-Higgs while they have non-zero 4-momentum, with
\begin{equation}
n^{i(q)}(\tau)=n^{0(q)}(\tau)\rightarrow\aleph_0
\end{equation}
leaving its 3-velocity be naturally observer-independent and equivalent to $c$.

\subsection{Eligible spin algebras}

Noticing that all rank-4 simple Lie algebras have positive definite Cartan algebras with metric $\delta_{\alpha\beta}=\text{diag}(+,+,+,+)$, we take the liberty of considering the quantized (tangent) frame as distributing on Cartan subalgebra $\{\mathbb{N}\cdot c_\alpha\}$ of a gauged rank-4 simple Lie algebra in which some of the root vectors in its root system are split to more than one generations of frame and Higgs.

A valid gauge selection is $spin(4,4)$ in Lie Group Cosmology (LGC) \cite{91,92}. Under the orthogonal regulation, we can find out the only eligible arrangement of $e_{\text{I}}^{\wedge}\phi_{\text{IV}}$, $e_{\text{I}}^{\vee}\phi_{\text{V}}$ there on $c_1$, and $e_{\text{II}}^{\wedge}\phi_{\text{IV}}$, $e_{\text{II}}^{\vee}\phi_{\text{V}}$ there on $c_2$, and $e_{\text{III}}^{\wedge}\phi_{\text{IV}}$, $e_{\text{III}}^{\vee}\phi_{\text{V}}$ there on $c_3$, and $e_{\text{O}}^{\wedge}\phi_{\text{IV}}^*$, $e_{\text{O}}^{\wedge}\phi_{\text{V}}$ there on $c_0$, leading to $dx^i\in\pm\sqrt{\mathbb{N}}$ and $dx^0\in\sqrt{\mathbb{N}}$ in SR, while $m\gtrless0$ only for massive particles $\mathtt{p}$ related to split $\phi_{\text{IV}(r_1)}$ and $m\lessgtr0$ only for $\phi_{\text{V}(r_2)}$, leaving nowhere for $e\phi_0$ and $e\phi_0^{\ *}$ therein. The arrangement is shown in Table \ref{tab:i}, where $\alpha, \beta, \gamma$, and centred $\varepsilon$ are the prime roots in Dynkin diagram of Lie algebra $d_4$, providing the $\sigma$ geometric system of $spin(4,4)$.
\begin{table}[!ht]
\centering
\begin{tabular}{*{6}{r}|*{6}{l}}
\textit{Axes:} & $\Sigma$ & $\omega_T$ & $\omega_S$ & U & $+$ & $-$ & $\Sigma$ & $\omega_T$ & $\omega_S$ & U & \\
\hline
$\alpha=$ & 0 & 1 & 1 & 0 & $e_{\text{I}}^\wedge\phi_{\text{IV}}$ &   & 0 & -1 & -1 & 0 & $e_{\text{I}}^\vee\phi_{\text{IV}}^*$\\
$\beta=$ & 1 & 0 & 0 & 1 & $e_{\text{II}}^\wedge\phi_{\text{IV}}$ &   & -1 & 0 & 0 & -1 & $e_{\text{II}}^\vee\phi_{\text{IV}}^*$\\
$\gamma=$ & 0 & -1 & 1 & 0 & $e_{\text{III}}^\wedge\phi_{\text{IV}}$ &   & 0 & 1 & -1 & 0 & $e_{\text{III}}^\vee\phi_{\text{IV}}^*$\\
$\varepsilon=$ & 0 & 0 & -1 & -1 & $e\phi_0^*$ &   & 0 & 0 & 1 & 1 & $e\phi_0$\\
$\alpha+\varepsilon=$ & 0 & 1 & 0 & -1 & $e_{\text{I}}^\wedge\phi_{\text{V}}^*$ &   & 0 & -1 & 0 & 1 & $e_{\text{I}}^\vee\phi_{\text{V}}$\\
$\beta+\varepsilon=$ & 1 & 0 & -1 & 0 & $e_{\text{II}}^\wedge\phi_{\text{V}}^*$ &   & -1 & 0 & 1 & 0 & $e_{\text{II}}^\vee\phi_{\text{V}}$\\
$\gamma+\varepsilon=$ & 0 & -1 & 0 & -1 & $e_{\text{III}}^\wedge\phi_{\text{V}}^*$ &   & 0 & 1 & 0 & 1 & $e_{\text{III}}^\vee\phi_{\text{V}}$\\
$\beta+\varepsilon+\gamma=$ & 1 & -1 & 0 & 0 & $\omega_\text{I}^\wedge$ &   & -1 & 1 & 0 & 0 & $\omega_\text{I}^\vee$\\
$\gamma+\varepsilon+\alpha=$ & 0 & 0 & 1 & -1 & $\omega_\text{II}^\wedge$ &   & 0 & 0 & -1 & 1 & $\omega_\text{II}^\vee$\\
$\alpha+\varepsilon+\beta=$ & 1 & 1 & 0 & 0 & $\omega_\text{III}^\wedge$ &   & -1 & -1 & 0 & 0 & $\omega_\text{III}^\vee$\\
$\alpha+\beta+\gamma+\varepsilon=$ & 1 & 0 & 1 & 0 & $e_{\text{O}}^\wedge\phi_\text{V}$ &   & -1 & 0 & -1 & 0 & $e_{\text{O}}^\vee\phi_\text{V}^*$\\
$\alpha+\beta+\gamma+2\varepsilon=$ & 1 & 0 & 0 & -1 & $e_\text{O}^\wedge\phi_\text{IV}^*$ &   & -1 & 0 & 0 & 1 & $e_\text{O}^\vee\phi_\text{IV}$
\end{tabular}
\caption{\label{tab:i} The arrangement on $\sigma$ system in frame-Higgs sector of $spin(4,4)$.}
\end{table}

\section{Conclusion}

The reliable basic structures of our 4-dimensional space $F_\geq\oplus F^3$ in SR and discrete tangent frame of quantized $e_{a(i)}$ on 4-dimensional Cartan base $\{\mathbb{N}\cdot c_\alpha\}$ are presented here.

Since $c^2\in\mathbb{N}>1$, we universally arranged the instant $(ds^\alpha)^2$, $(dx^\alpha)^2$, $(V^\alpha)^2$, $(U^\alpha)^2$, $(v^i)^2$, $(a^\alpha)^2$, $(n^\alpha e_{a(i)})^2$, and $(d^\alpha e_{a(i)})^2$ in $\mathbb{N}$ of every observed massive particle $\mathtt{p}_{(q)}$. If a rest frame of $\mathtt{p}_{(q)}$ is selected, it turns to Eq.$(4.39)$. The constant $k^2=(c^2-1)!$ in $\mathbb{N}$ is of a continuous approximation $t_0^{\ 0}(x)=-c\cdot t_i^{\ i}(x)$ (no sum) of the constant, diagonal invertible embedder $t_\alpha^{\ \beta}(x)$ for $dx^\beta=ds^\alpha \bar{t}_\alpha^{\ \beta}$. In this case, the constant $c$ is declared from even distributions of massless, zero-momentum, split particles of the quantized frame $e_{a(i)}$ with their positions $n^\alpha(e_{a(i)})\in\mathbb{N}$ on $c_\alpha$. As well as other massive particles $\mathtt{p}$, the combined frame-Higgs $e_a\phi_{\text{IV}(s)}$ are bosons related to some split frame $e_{a(i)}$ and Higgs $\phi_{\text{IV}(r)}$ from generators of a larger gauge algebra, receiving its unique 3-velocity, 4-velocity, and rest mass at each $\tau\in\mathbb{N}$.

Within the evolution of proper time $\tau$, its CM-coordinate $\{x^{\alpha(s)}\}$ and our history live in the domains $S_\alpha$ of $S_i\subset F$ and $S_0\subset F_\geq$.

\acknowledgments

The author would like to thank Jian Guo, Wen-tao Ren, Chuang-hui Ye, Ai-guo Huang, Qiang Li, and Zi-ping Zhu for advice, and Almer Wang, Zhi-ren Yv, SRT, Yi-fu Chen, Hao-lin Liu, Tony WED, Ren-XVIII for encouragements. I would also like to thank Sheldon Smith, Garrett Lisi and Justin Vines for helpful discussion. This album is supported by the Principal Office of Guangdong Olympic School.

\paragraph{Note added.}This paper lives in 7-2-3 Mini Album: $e\phi\ 002$.

\end{document}